\documentclass[prl,aps,twocolumn,superscriptaddress]{revtex4}
\usepackage{graphicx,graphics}
\usepackage{dcolumn}				
\usepackage{amssymb}
\usepackage{hyperref}

\newcommand{\beqa}{\begin{eqnarray}}
\newcommand{\eeqa}{\end{eqnarray}}
\newcommand{\beq}{\begin{equation}}
\newcommand{\eeq}{\end{equation}}

\begin{document}
\title{Dynamics of the entanglement between two oscillators in the same environment}

\author{Juan Pablo \surname{Paz}}
\affiliation{Departamento de F\'{\i}sica, FCEyN, UBA, Pabell\'on 1,
Ciudad Universitaria, 1428 Buenos Aires,
 Argentina}

\author{Augusto J. \surname{Roncaglia}}
\affiliation{Departamento de F\'{\i}sica, FCEyN, UBA, Pabell\'on 1,
Ciudad Universitaria, 1428 Buenos Aires,
Argentina}

\begin{abstract}
We provide a complete characterization of the evolution of entanglement 
between two oscillators coupled to a common environment. For initial Gaussian
states we identify three phases with different qualitative long time behavior:
There is a phase where entanglement undergoes a sudden death (SD). Another
phase (SDR) is characterized by an infinite sequence of events of sudden 
death and revival of entanglement. In the third phase (NSD) there is 
no sudden death of entanglement, which persist for long time. The phase 
diagram is described and analytic expressions for the boundary between 
phases are obtained. Numerical simulations show the accuracy of 
the analytic expressions. These results are applicable to a large 
variety of non--Markovian environments. The case of non--resonant 
oscillators is also numerically investigated. 
\end{abstract}

\maketitle

Entanglement is responsible for the most counterintuitive aspects of quantum 
mechanics. In the early days of quantum physics, entanglement 
motivated many philosofical discussions but nowadays it is viewed merely as a
source of surprises but, mainly, as a physical resource. In fact, entanglement 
is the key ingredient for quantum teleportation and is believed to be the origin of 
the power of quantum computers. Systems of qubits interacting with common 
or independent environment where extensively
analyzed in the literature \cite*{Braun02,Benatti03,Oh06,Anastopoulos06}. In such context, the 
existence of a peculiar property was noticed in \cite{Yu04}: Entanglement, as 
opposed to energy, can disappear from a system in a finite time 
(this phenomenon was denoted as "sudden death" SD of entanglement).

Here we study the evolution of the entanglement between two oscillators 
interacting with a common bosonic environment. Entanglement dynamics for 
such kind of systems was analyzed before and a variety of results is available
in the literature. For example: It was shown that the environment
may completely dissentangle initially entangled state (a two--mode squeezed state). 
Thus, in  \cite{Paris02,Serafini04,Olivares07} the existence of sudden death 
was proved and analyzed for such case. The evolution 
of the same type of initially entangled state was analyzed when the two 
oscillators interact with a common bath under a simplifying assumption: the Markovian 
approximation \cite{Prauzner04,Dodd04,Benatti06-2}. In that case, a condition for the 
existence of an asymptotic entangled state was deduced \cite{Prauzner04}. If such 
condition is not satisfied (see below) sudden death takes place. 
Also, it was shown that the interaction with a common environment opens the 
door to the creation of an entangled state from an initial separable 
state \cite{Benatti06-2}. Most of these works were done under some sort of Markovian
approximation. More recently, the non-Markovian regime  was 
analyzed \cite{Liu07,An07,Horhammer07}. Under special conditions a simple 
result emerged: The final state remains entangled but half of the initial 
entanglement is lost \cite{An07}. 

Our work provides a unified picture to understand the origin of 
the different qualitative behaviors (``phases'') that characterize the evolution 
of entanglement for long times. We will fully characterize 
these phases of the evolution and obtain equations describing the boundary between them. 
For this purpose we will use a well known tool: the exact master equation for quantum 
Brownian motion \cite{HuPazZha92}. We consider two harmonic oscillators ($x_1$ and $x_2$) 
bilinearly coupled between them and with a collection of harmonic 
oscillators \cite{FeyVer63,CalLeg83,HuPazZha92,Chou07}. The total Hamiltonian is 
$H=H_S+H_{int}+H_{env}$ where  
\beqa
H_S&=&\frac{p_1^2+p_2^2}{2m}+\frac{1}{2}m(\omega_1^2 x_1^2+\omega_2^2 x_2^2)+mc_{12} 
x_1 x_2 \label{eq:H_ent} \nonumber \\
H_{env}&=&\sum_{n=1}^{N}(\frac{\pi_n^2}{2m_n}+\frac{m_n}{2} w_n^2 q_n^2),\\
H_{int}&=&(x_1+x_2)\sum_{i=1}^{N}c_n q_n.\nonumber
\eeqa
Using coordinates $x_\pm=(x_1\pm x_2)/\sqrt{2}$, the Hamiltonian transforms into 
$H_S=(p_+^2+p_-^2)/2m + m(\omega_-^2 x_-^2 +\omega_+^2 x_+^2)/2+m c_{+-} x_+x_- $, 
where 
$\omega_{\pm }^2=(\omega_1^2+\omega_2^2)/2\pm c_{12}$ and $c_{+- }=(\omega_1^2-\omega_2^2)/2$. 
This model can be exactly solved assuming that the initial state of the environment is 
thermal with initial temperature $T$ \cite{HuPazZha92}, and entanglement properties between the system and
the environment were studied in \cite{Eisert02}.
The resonant case ($\omega_1=\omega_2$), where the $x_\pm$ oscillators are decoupled, is the simplest: The 
exact master equation for the reduced density matrix of the two 
oscillators $\rho$ is ($\hbar=1$) \cite{HuPazZha92,Chou07}:
\beqa
\dot\rho&=&{1\over{i}}[H_R,\rho]-i\gamma(t)[x_+,\{p_+,\rho\}]-\nonumber\\
&-&D(t)[x_+,[x_+,\rho]]-f(t)[x_+,[p_+,\rho]].\label{eq:mastereq}
\eeqa
Here, the renormalized Hamiltonian is $H_R=H_S+m\delta\omega^2(t)x_+^2/2$. The 
coefficients $\delta\omega^2(t)$, $\gamma(t)$, $D(t)$ and $f(t)$ depend on the 
spectral density of the environment, defined as  $J(\omega)=\sum_n c_n^2\delta(\omega-w_n)/2m_n  w_n$, 
and also on its  initial temperature. The explicit form of these coefficients is rather cumbersome  
and was studied in detail elsewhere \cite{HuPazZha92,Fleming07}. We will focus on the  
ohmic environment with a spectral density $J(\omega)=2m\gamma_0\omega \theta(\omega-\Lambda)/\pi$. 
The high frequency cutoff $\Lambda$ defines a characteristic timescale 
$\Lambda^{-1}$ over which the coefficients $\gamma(t)$ and
$\delta\omega^2(t)$ vary. For times $t\gg\Lambda^{-1}$ these two temperature independent 
coefficients settle into asymptotic values: $\gamma(t)\rightarrow\gamma=2\gamma_0$ 
and $\delta\omega^2(t)\rightarrow -4m\Lambda\gamma/\pi$. The time dependent 
frequencies $\Omega^2_{1,2}(t)=\omega^2_{1,2}+\delta\omega^2(t)/2$ approach cutoff independent
values only if the bare frequencies $\omega_{1,2}$ have an appropriate dependence on the cutoff. 
The coupling constant $c_{1,2}$ must also be renormalized in the same way so that the time 
dependent coupling $C_{12}(t)=c_{12}+\delta\omega^2(t)/2$ approaches a finite cutoff 
independent value. 
The behavior of the diffusion 
coefficients $D(t)$ and $f(t)$ is more complicated and depend on the initial temperature. 
For the moment we just need to mention here that for realistic environments 
these coefficients approach asymptotic values after a temperature-dependent 
time (which equals $\Lambda^{-1}$ only in the high temperature regime). 
A word on notation: upper case letters will be used to denote renormalized quantities. 
The time label will be omitted when referring to the asymptotic value of 
the corresponding function (i.e., $\Omega_{1,2}$ denotes the asymptotic value 
of the renormalized frequency of the oscillators, etc). 

From the master equation (\ref{eq:mastereq}) we can derive simple evolution 
equations for the covariance matrix $V_{ij}(t)=Tr(\rho(t)\{r_i,r_j\})/2-Tr(\rho(t)r_i)Tr(\rho(t)r_j)$ 
where $i,j=1,\ldots,4$ and $\vec r=(x_-,p_-,x_+,p_+)$. Some of these equations are particularly illuminating: 
In fact, equations for the covariances split into two blocks of $2\times 2$. The evolution of the first block 
formed with the second moments of $x_-$ and $p_-$ corresponds to a free oscillator with frequency 
$\Omega^2_-(t)$. The evolution of the second block, formed with the second moments of $x_+$ and $p_+$,
satisfys:
\beq
{d\over{dt}}
\left({\langle p_+^2 \rangle\over{2m}} + 
      \frac{m}{2} \Omega^2(t)\langle x_+^2 \rangle \right)=
-\frac{2\gamma(t)}{m}\langle p_+^2\rangle+{D(t)\over{m}}, 
\eeq
\beq
{1\over 2}\frac{d^2\langle x_+^2\rangle}{dt^2}
+\gamma(t)\frac{d\langle x_+^2\rangle}{dt}
+\Omega^2(t)\langle x_+^2\rangle = 
\frac{\langle p_+^2\rangle}{m^2}-\frac{f(t)}{m}. \nonumber\label{eq:momenta}
\eeq
These equations contain most of the necessary information to fully analyze the evolution of the 
entanglement between initial Gaussian states. To solve them exactly we need to know the time 
dependent coefficients that appear in the master equation. 
But, remarkably enough,  we can use the above equations to understand the 
qualitative behavior of entanglement. For this, we only need to assume that the 
time dependent coefficients in (\ref{eq:mastereq}) approach asymptotic values. In such 
case, there is a stable stationary solution where the dispersions 
$\Delta^2 x_+=\langle x_+^2\rangle$ and $\Delta^2 p_+=\langle p^2_+\rangle$ are 
\beq
\Delta p_+= \sqrt{{D \over 2 \gamma}}; \quad \Omega \Delta x_+= \sqrt{{D \over 2 m^2\gamma} - 
{f \over m}}. \label{eq:Dandf}
\eeq

Entanglement for Gaussian states is entirely determined by the properties of the covariance 
matrix $V_{ij}$. In fact, a good measure of entanglement for such states is the 
logarithmic negativity  $E_{\mathcal N}$ \cite{Vidal02,PHdEisert}, computable 
as \cite{Vidal02,PHdEisert,Adesso04}: 
\beq
E_{\mathcal N}=\max\{0,-\ln(2\nu_{\min})\},
\eeq
where $\nu_{\rm min}$ is the smallest symplectic eigenvalue of the partially transposed 
covariance matrix. There are known expressions for $E_{\mathcal N}$ for particularly 
relevant Gaussian states which will be used as initial conditions in our study. For this
reason it is useful to mention them here: For the two-mode squeezed state, obtained 
from the vacuum by acting with the operator $\exp(-r(a_1^\dagger a_2^\dagger- a_1 a_2))$,  
we have $E_{\mathcal N}=2|r|$. For this state the dispersions satisfy the minimum 
uncertainty condition $\delta x_+ \delta p_+ =\delta x_- \delta p_- =1/2$. The squeezing 
factor determines the ratio between variances since 
$m\Omega\delta x_+ /\delta p_+ =\delta p_- /(m\Omega\delta x_-) =\exp(2r)$. 
As $r\rightarrow\infty$ the state becomes localized 
in the $p_+$ and $x_-$ variables approaching an ideal EPR state \cite{EPR}. 
Another initial state we will consider is a separable squeezed state for which 
$m\Omega\delta x_{1,2}/\delta p_{1,2}=\exp(2r)$.

The evolution of entanglement for initial Gaussian states can then be simply 
analyzed by using the previous results. Let us consider a general initial state. 
For the resonant case, the $x_-$ virtual oscillator decouples exactly. Using 
the equations (\ref{eq:momenta}) we see that $\Delta x_+$ and $\Delta p_+$ 
approach asymptotic values after a timescale $1/\gamma$, fixed by the dissipation 
rate. After that time, the covariance matrix in the $(x_+, x_-)$ bases has a $2\times 2$ 
block with oscillatory functions (corresponding to the $x_-$ oscillator) and another 
$2\times 2$ block, corresponding to the $x_+$ virtual oscillator, which is diagonal. 
Using this form for the covariance matrix (and changing basis to obtain covariances 
of the original $x_{1,2}$ oscillators) it is simple to find its smallest 
symplectic eigenvalue and compute the following logarithmic negativity  \cite{generalR} 
\beq
E_{\mathcal N}(t)\rightarrow\max\{0,\tilde E_{\mathcal N}+ \Delta E_{\mathcal N} G(t)\}. 
\eeq
Where $G(t)$ is a function with period $\pi/\Omega_-$ in $\{-1,1\}$.
The mean value $\tilde E_{\mathcal N}$ about which $E_{\mathcal N}$ oscillates is
\beq
\tilde E_{\mathcal N}=\max\{r,r_{crit}\} -{1\over 2} \ln(2 \Delta x_+ \Delta p_+),
\eeq
and the amplitude of the oscillation is:
\beq
\Delta E_{\mathcal N}=\min\{r,r_{crit}\}.\nonumber
\eeq
Here $r$ is the initial squeezing factor defined as 
\mbox{$r=|{1 \over 2}\ln(m \Omega_- {\delta x_- \over \delta p_-})|$}, 
$r_{crit}$ is related to the squeezing factor of the equilibrium state for the  $x_+$-oscillator defined 
as \mbox{$r_{crit}=|{1 \over 2}\ln(m \Omega_- {\Delta x_+ \over \Delta p_+})|$}. 
The dispersions $\Delta x_+$ and $\Delta p_+$ are the asymptotic 
values (\ref{eq:Dandf}). 

These simple results will enable us to draw general conclusions about
the dynamics of entanglement for long times. Three qualitatively different 
scenarios emerge: First, entanglement may persist for arbitrary long times. 
This phase, which we call ``NSD'' (for no-sudden death), is realized when 
the initial state is such that $\tilde E_{\mathcal N}-\Delta E_{\mathcal N}>0$, 
which translates into $|r-r_{crit}|>\ln(2\Delta x_+\Delta p_+)/2=S_r$. 
Then, there is a phase where entanglement undergoes an infinite sequence of events 
of ``sudden death'' and  ``sudden revival'' \cite{Yonac06,Yonac07}. This occurs if the initial state is
such that $|\ln(1/2m\Omega_-\Delta^2 x_+)/2|\le r\le \ln(2\Delta^2 p_+/m\Omega_-)/2$. We 
denote this phase as ``SDR'' (for sudden death and revival). Finally, a third  
phase characterized by a final event of ``sudden death'' of entanglement may 
be realized if $r\le -E_c\equiv-\ln(1/2m\Omega_-\Delta^2x_+)/2$. This 
phase is simply denoted as ``SD'' (for sudden death). 

\begin{figure}
\includegraphics[width=8.7cm]{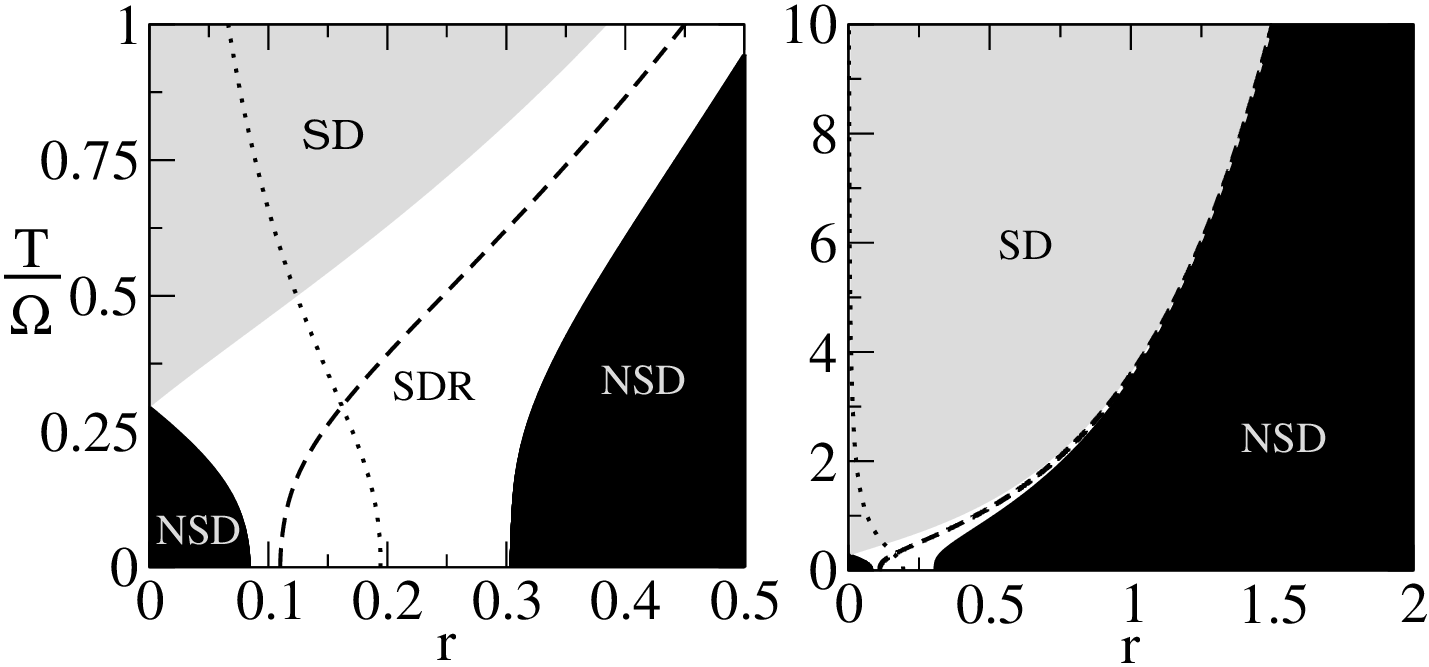}
\caption{Phase diagram for ohmic environment ($\Omega=1$, $\gamma_0=0.15$, $\Lambda=20$, $m=1$, $C_{12}=0$). 
The sudden death (SD), no-sudden death (NSD) and sudden death and revival (SDR) phases 
describe the three different qualitative long time behaviors for the entanglement 
between two oscillators interacting with the same environment. The SDR phase is centered 
about the dashed line $S_r$ and has a width given by the dotted line $r_{crit}$. 
This is the case for temperatures above the one for which $S_r=r_{crit}$. Below this temperature
the role of $S_r$ and $r_{crit}$ are interchanged. SDR separates the SD and 
NSD phases. The low temperature NSD island is due to non--Markovian and 
non--perturbative effects. $\tilde E_{\mathcal N}$  in the NSD phase is the distance
to the dashed line for $r > r_{crit}$, and the distance between the  dashed and dotted lines for $r\leq r_{crit}$.
} 
\label{fig:phases}
\end{figure}

Depending on the properties of the environment (initial temperature, damping rate, etc)
a given initial state (parameterized by the squeezing $r$) will belong to one of the 
three phases. For the ohmic environment we can use exact analytic expressions for the
coefficients $D$ and $f$ \cite{Fleming07} to obtain the phase diagram displayed in 
Fig. \ref{fig:phases}. 
In the diagram, the areas corresponding to each of the three phases are displayed. As a reference,
we also include the curves that show the temperature dependence of $S_r$
and $r_{crit}$ (dashed and dotted lines respectively).

The phase diagram provides complete information about the asymptotic behavior of 
entanglement. It is worth mentioning some features of the 
diagram. The NSD phase present at low temperatures is purely non--Markovian
and non--perturbative. Its area shrinks as the damping rate decreases. This phase includes
the coherent states, that can become entangled at very low temperature. 
The zero temperature line is also interesting: It contains states in the NSD phase 
for small and large squeezings. However for an intermediate range of squeezings 
centered about $r_{crit}$ states belong to the SDR phase. 

To the contrary, the high temperature region of the diagram is quite different. Thus, 
for high temperatures we have $E_c<0$ (which implies that coherent states do
not get entangled) and also $r_{crit}\rightarrow 0$ (which implies that the region
covered by the SDR phase becomes narrower). Thus, for high temperatures 
initial states with large squeezing 
($r>\ln(2\Delta x_+\Delta p_+)/2=S_r$) retain some of their entanglement while those 
with squeezing factors smaller than the critical value $S_r$ suffer from sudden death.
We remark that this is almost the same condition obtained previously for persistence 
of entanglement in the Markovian regime \cite{Prauzner04}. However, our analysis shows 
that the boundary
between SD and NSD phases is rather subtle: for any finite temperature 
the two phases are separated by a very narrow portion of SDR phase (in this phase there 
are oscillations of the entanglement whose amplitude, $r_{crit}$ 
decreases as temperature grows). This is yet another interesting non-Markovian effect 
identified by our analysis. A final comment on the phase diagram: The NSD phase is 
characterized by a non-vanishing 
asymptotic entanglement that can be quantified in a straightforward way from the 
phase diagram itself. The average value of the logarithmic negativity is simply the distance to 
the dashed line (which signals the midpoint of the SDR phase) or just the distance between the 
dashed and dotted lines for $r\leq r_{crit}$.
\begin{figure}
\includegraphics[width=8.8cm]{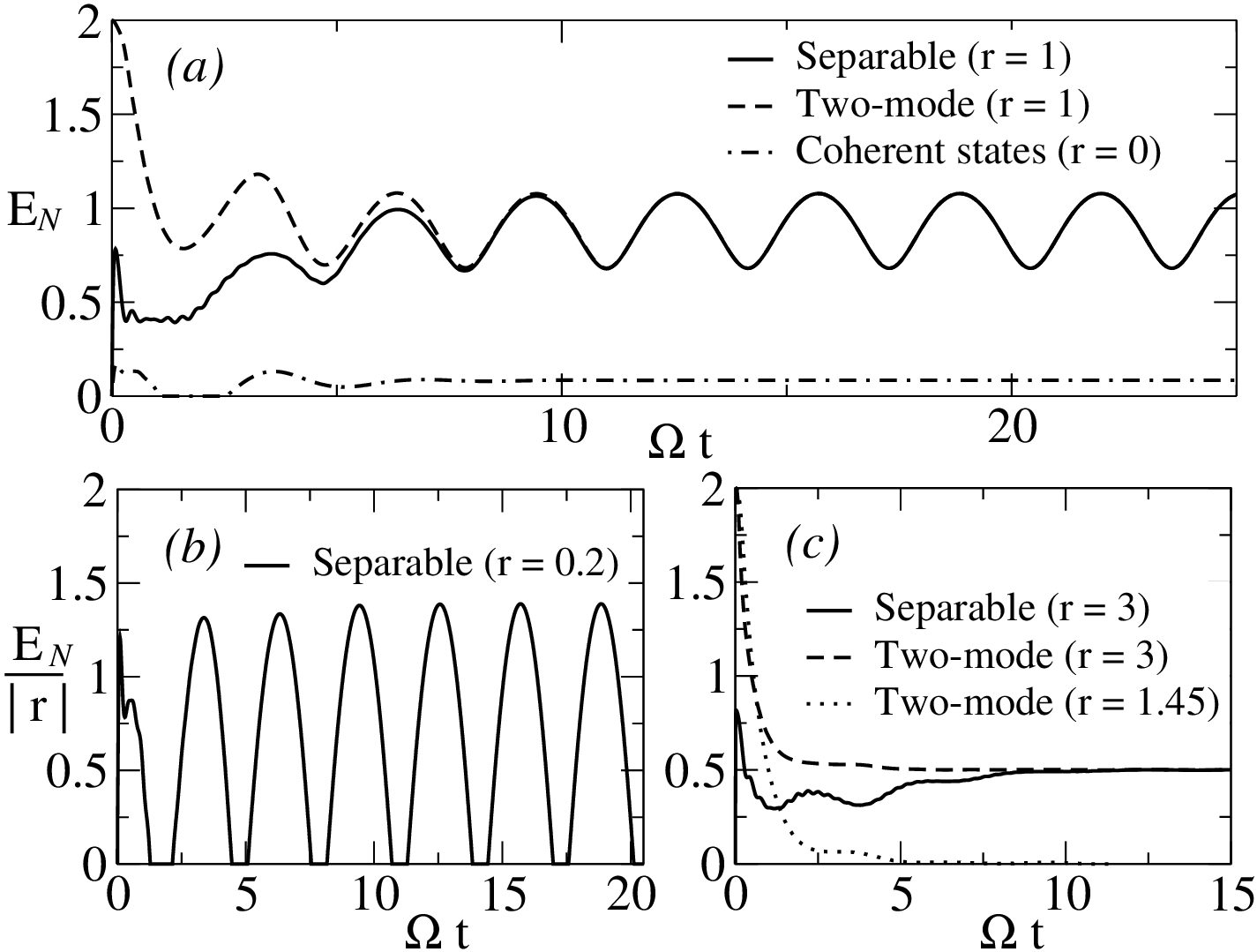}
\caption{Logarithmic negativity for resonant oscillators 
in the same environment. $(a)$ For $T=0$ the NSD phase appears both for  
for large and small squeezing. Asymptotic behavior of initially entangled or 
separable states only depends on $r$. The amplitude of oscillations vanishes 
when $r\rightarrow 0$.
$(b)$ The SDR phase appears for intermediate values of squeezing at zero temperature.
$(c)$ $T/\Omega=10$, the SD phase appears for small $r$ and NSD phase for large squeezings, oscillations 
in the steady state are attenuated as the temperature increases.}
\label{fig:entTime}
\end{figure}

The validity of the above rather simple but nontrivial analytical predictions can be verified by 
an exact numerical solution to the problem. We considered an ohmic environment with the 
following parameters: $\gamma_0=0.15$, $\Lambda=20$, $m=1$, $\Omega=1$, $C_{12}=0$ (extension to the case where the 
oscillators interact can be easily done). The
results for the evolution of $E_{\mathcal N}$ are shown in Fig. \ref{fig:entTime}. 
We clearly see that the final entanglement produced by different initial states only depends 
upon the squeezing factor $r$. The existence of events of sudden death (and revivals) can 
also be seen 
from the numerical solution, and is predicted from our analytic results characterizing
the SDR phase, Fig. \ref{fig:entTime}. It is also noticeable that the amplitude of the 
oscillations 
vanishes when $f/D\rightarrow 0$, which is a feature of the high temperature limit. 
The oscillating behavior of the final entanglement is a consequence of the non-zero 
asymptotic value of the 
anomalous diffusion coefficient $f(t)$, (see Fig. \ref{fig:entTime}). Moreover, this same 
coefficient is also responsible for 
the generation of asymptotic entanglement in initial coherent states. 
But in this case the entanglement is constant $E_\mathcal{N}=E_c$ and increases with $\gamma_0$, 
as the final state of the oscillator $x_+$ becomes more and more localized in position.
This is another  purely non--Markovian effect that is entirely lost in 
the RWA-Markovian approximation \cite{Prauzner04}. 
With the numerical solution, we verified that every possible temporal 
evolution that can be obtained by changing parameters fits either in the 
NSD, SDR or the SD phase. The boundaries between phases is
also accurately described by the conditions mentioned above. 
 
\begin{figure}
\vspace{0.1cm}
\includegraphics[width=8.8cm]{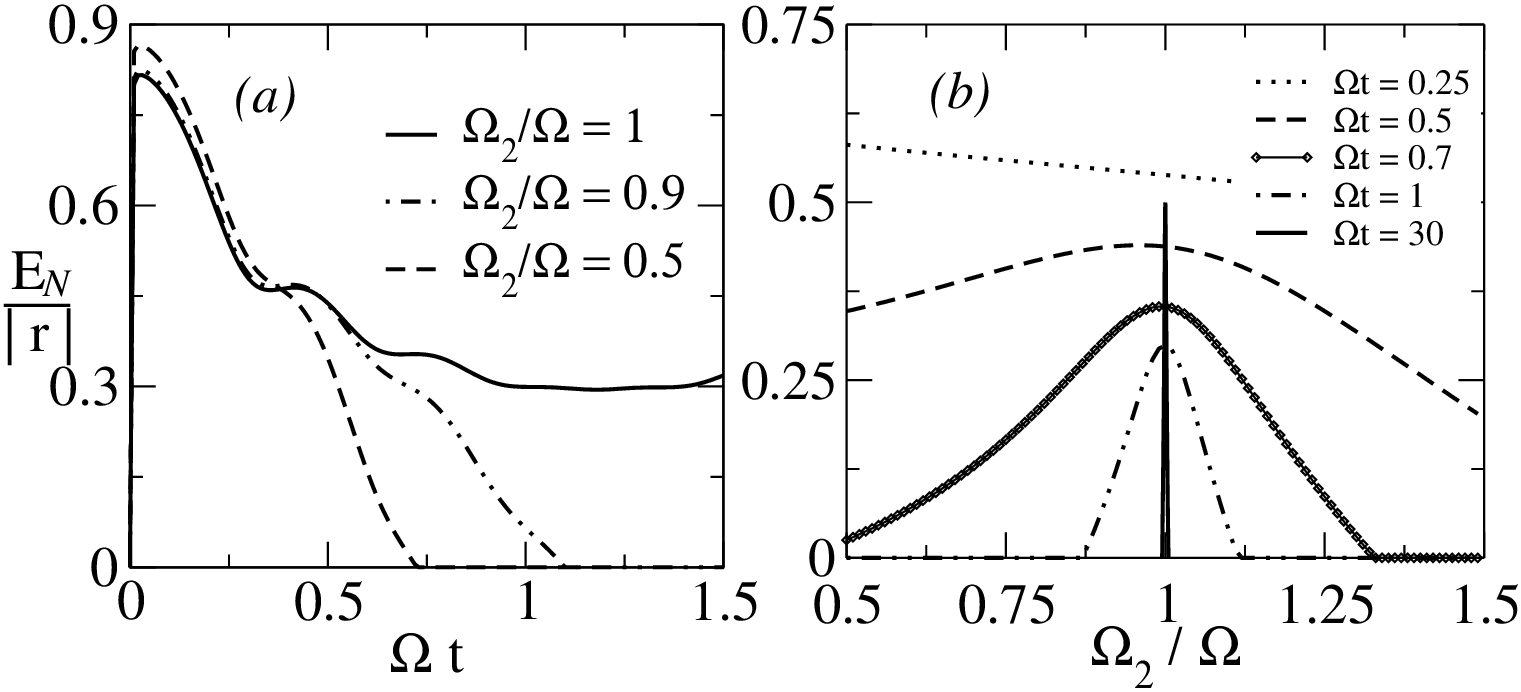}
\caption{Entanglement dynamics for non-resonant oscillators initially in a squeezed separable state ($r=3$) and
$T/\Omega=10$. $(a)$ Entanglement is created between non-resonant oscillators but 
it banishes in finite time. $(b)$ $E_\mathcal{N}$ for different times as a function
of the frequency of the second oscillator, the resonant condition is essential for asymptotic entanglement.} 
\label{fig:entFreq}
\end{figure}

The above properties are valid under a single important assumption: 
the two oscillators are resonant. If this is not the case the analysis becomes more 
complicated: the master equation is no longer valid since there are terms coupling 
the $x_-$ and $x_+$ virtual oscillators. Due to this coupling the $x_-$ oscillator is  
not protected from the environment and also reaches an equilibrium state. In 
Fig. \ref{fig:entFreq} we show how things change when we move away from the resonance 
condition. It is clear that $E_{\mathcal N}$ decays much faster in this case. Also, the value 
of $E_{\mathcal N}$ was analyzed away from resonance for different times. 
The result is shown in Fig. \ref{fig:entFreq} where it is clear that the resonance
peak becomes sharper and sharper as time grows. 

Our results provide a complete picture of the possible behavior of entanglement for  
resonant oscillators coupled to a common environment. The same analysis is 
valid for any environment such that an asymptotic value of the coefficients of the
master equation is obtained. For example, for super--ohmic environments the damping 
coefficient approaches a small value while both $f(t)$ and $D(t)$ vanish asymptotically. 
Our analysis  predicts that there will not be an equilibrium 
value for the dispersions of the $x_+$ oscillator and that entanglement
will oscillate approximately recovering its initial value after some time. 
This is indeed observed in our
numerical simulations (contrary to what is indicated in \cite{An07}, where the 
results obtained seem
to be an artifact produced by an unreasonably low cutoff). 
Finally, we remark once again that our results for the asymptotic entanglement are 
highly dependent on the validity of
the resonance condition. For non-resonant oscillators the generic fate for long times is 
disentanglement due to the interaction with the environment. This issue will be analyzed in detail
elsewhere. Authors acknowledge support from CONICET and  
Anpcyt (Argentina). 


\begin{thebibliography}{27}
\bibitem[{gen()}]{generalR}
\bibinfo{note}{Here we consider the case where $\delta x_- \delta p_-=1/2$. In
  general, we can apply the same results replacing $\Delta x_+\Delta p_+
  \rightarrow 2 \Delta x_+\Delta p_+\delta x_- \delta p_-$.}
\end{thebibliography}

\end{document}